\documentclass[onecolumn,12pt]{article}
\pdfoutput=1
\usepackage{jheppub}

\usepackage{tikz}

\usepackage[font={small}]{caption}
\usepackage{subfigure}
\usepackage{graphicx}
\usepackage{dcolumn}
\usepackage{float}
\usepackage[normalem]{ulem}

\usepackage{mathrsfs} 

\usepackage{multicol}
\usepackage{cancel}
\usepackage{mathtools}
\usepackage{hyperref}

\usepackage{subfigure}
\usepackage{xcolor}
\def\({\left(} \def\){\right)}
\def\[{\left[} \def\]{\right]}

\newcommand{\bit}{\begin{itemize}}
	\newcommand{\eit}{\end{itemize}}

\def\a{\alpha}

\def\b{\beta}

\def\l{\lambda}

\def\s{\sigma}

\def\le{\left(}
\def\ri{\right)}

\newcommand{\be}{\begin{equation}}
\newcommand{\ee}{\end{equation}}
\newcommand{\bea}{\begin{eqnarray}}
\newcommand{\eea}{\end{eqnarray}}

{\left\lbrace\begin{array}{@{}l@{}}}%
{\end{array}\right.}

\renewcommand{\eqref}[1]{(\ref{#1})}

\begin{document}

\begin{flushright}
	\hfill{OU-HET-1170}
\end{flushright}

 \author{Takanori Anegawa,}
 \author{Norihiro Iizuka,}
 \author{Sunil Kumar Sake,}
 \author{Nicol\`o Zenoni}
 \affiliation{\it Department of Physics, Osaka University, 
Toyonaka, Osaka 560-0043, JAPAN}

\emailAdd{takanegawa@gmail.com}
\emailAdd{iizuka@phys.sci.osaka-u.ac.jp}
\emailAdd{sunilsake1@gmail.com}
\emailAdd{nicolo@het.phys.sci.osaka-u.ac.jp}

\vspace{1cm}

\abstract{Volume complexity in dS$_2$  remains $O(1)$ up to a critical time, after which it  suddenly diverges. On the other hand, for the dS$_2$ solution in JT gravity,  there is a linear dilaton which smoothly grows towards the future infinity. From the dimensional reduction viewpoint, the growth of the dilaton is due to the expansion of the orthogonal sphere in higher-dimensional dS$_d$ ($d \ge 3$). 
Since in higher dimensions complexity becomes very large even before the critical time, by properly taking into account the dilaton, the same behavior is expected for complexity in dS$_2$ JT gravity.
We show that this expectation is met by the complexity = action (CA) conjecture. For this purpose, we obtain an appropriate action for dS$_2$ in JT gravity, by dimensional reduction from dS$_3$. In addition, we discuss complexity = ``refined volume'' where we choose an appropriate Weyl field-redefinition such that refined volume avoids the discontinuous jump in time evolution. }

\title{Is Action Complexity better for de Sitter space in Jackiw-Teitelboim gravity?} 

\maketitle

\setcounter{tocdepth}{2}
\section{Introduction}

The AdS/CFT correspondence \cite{Maldacena:1997re} has been very successful and has led to remarkable insights into the bulk gravity theory. Anti-de Sitter (AdS) spacetime effectively acts  as a box: bulk quantities allow for a boundary definition, and computations on both sides match.
Nevertheless, our universe rather resembles de Sitter (dS) spacetime.
Despite the significant success of the AdS/CFT, the situation becomes complicated when we simply change the sign of the cosmological constant to obtain dS spacetime. In this case, no natural notion of a timelike boundary exists where a conventional dual quantum theory can be located.   
To circumvent the issue, many suggestions on where the dual observer should live have been put forward. A possibility is to consider the spacelike future infinity of dS spacetime \cite{Strominger:2001pn,Bousso:2001mw,Balasubramanian:2002zh}. A different approach makes instead use of the stretched horizon, a timelike hypersurface close to the cosmological horizon in the dS static patch \cite{Susskind:2021omt}.
For two-dimensional dS, the dual theory placed on the stretched horizon has been proposed to be a high-temperature limit of the double-scaled Sachdev-Ye-Kitaev (SYK) model with  $q\sim \sqrt{N}$, where $q$ specifies the size of the interaction terms in the Hamiltonian schematically as $H\sim \psi^q$ and $N$ is the number of flavours of fermions \cite{Susskind:2021esx}. Some results on this model can be found in \cite{Susskind:2022dfz,Bhattacharjee:2022ave,Lin:2022nss,Susskind:2022bia,Rahman:2022jsf}.

On the other hand,  
two-dimensional 
Jackiw-Teitelboim (JT) gravity \cite{Jackiw:1984je,Teitelboim:1983ux} 
has been actively studied in recent years. JT gravity in AdS background, which is an effective low-energy theory for near-extremal Reissner-Nordstr\"om black holes in four dimensions, has deep connections with the SYK \cite{Maldacena:2016hyu,Sarosi:2017ykf} and the random matrix theory \cite{Turiaci:2020fjj}. 
The reason for our interest in JT gravity arises from its simplicity and analytic tractability, as well as from the hope that it may be regarded as a solvable toy model for higher-dimensional quantum gravity. Given this, it is quite natural to investigate JT gravity in the dS background too.   

Before we proceed, we comment on the connection to quantum information.  
An interesting question would be to understand how the bulk spacetime emerges from the boundary degrees of freedom, to which purpose quantum information may play a central role. 
As a remarkable example in the AdS/CFT framework, 
the Ryu-Takayanagi (RT) prescription relates entanglement entropy in a CFT to the area of extremal codimension-two bulk surfaces anchored at the AdS boundary \cite{Ryu:2006bv,Ryu:2006ef, Hubeny:2007xt}.\footnote{Generalization of the RT proposal to dS spacetime has been discussed in \cite{Susskind:2021esx,Shaghoulian:2021cef,Shaghoulian:2022fop}, where the RT surface has been conjectured to be anchored at the dS stretched horizon.} 
However, 
in terms of the dual theory in AdS, 
entanglement entropy does not suffice \cite{Hartman:2013qma} to capture the growth of the Einstein-Rosen bridge inside a black hole, which continues even  after the thermalization time.  
The appropriate dual quantity to be considered is quantum complexity, which keeps growing for an exponentially large time \cite{Susskind:2014moa,Susskind:2014rva}.
In the context of computer science, where complexity is identified with the minimum number of elementary gates required to build a target state from a reference one, this property has been shown to hold for random quantum circuits \cite{Haferkamp:2021uxo}.
A geometric definition of complexity as the length of shortest geodesics on the manifold of unitary operators has been introduced in \cite{Nielsen} for quantum systems with a finite number of degrees of freedom. 
Chaotic behaviour of complexity geometry, which is expected to reflect the black hole interior evolution, has been found to be related to negative curvature of the manifold of unitary operators \cite{Brown:2016wib,Brown:2019whu,Auzzi:2020idm,Brown:2021euk,Basteiro:2021ene}. Applications of the geometric approach to free quantum field theories have been studied in \cite{Jefferson:2017sdb,Chapman:2017rqy,Hackl:2018ptj,Khan:2018rzm}. Extensions of complexity to CFTs have been addressed in \cite{Caputa:2017urj,Caputa:2018kdj,Belin:2018bpg,Erdmenger:2020sup,Flory:2020eot,Flory:2020dja,Chagnet:2021uvi,Erdmenger:2021wzc}.
See \cite{Susskind:2014moa,Susskind:2018pmk,Chapman:2021jbh} for reviews on complexity.

Three main conjectures have been put forward for the holographic dual of complexity: complexity = volume (CV) \cite{Stanford:2014jda}, relating complexity to the volume of codimension-one extremal surfaces, complexity = action (CA) \cite{Brown:2015bva,Brown:2015lvg}, invoking the full gravitational action of the Wheeler-DeWitt (WDW) patch, and CV 2.0 \cite{Couch:2016exn}, involving the spacetime volume of the WDW patch. 
Recently, it has been pointed out that an infinite class of observables defined on codimension-one \cite{Belin:2021bga} and codimension-zero  \cite{Belin:2022xmt} bulk regions exist which grow linearly at late times in the classical regime and are subject to the switchback effect   \cite{Susskind:2014jwa,Susskind:2018pmk}.
Such arbitrariness in selecting the bulk counterpart of complexity may mirror the intrinsic ambiguity in the choice of elementary gates in the complexity definition. 
Holographic complexity for AdS in JT gravity and in other JT-like two-dimensional dilaton-gravities has been investigated in \cite{Brown:2018bms,Goto:2018iay, Alishahiha:2018swh,Jian:2020qpp,Cai:2020wpc, An:2018xhv,Akhavan:2018wla}.
{Semi-classical corrections give a volume that saturate at late times \cite{Iliesiu:2021ari,Alishahiha:2022kzc,Alishahiha:2022exn}.}
In recent years, holographic complexity calculations have also been performed
in hybrid spacetimes interpolating between AdS and dS \citep{Chapman:2021eyy,Auzzi:2023qbm}, for general-dimensional dS spacetime in Einstein gravity \cite{Jorstad:2022mls}, and for two-dimensional dS spacetime in JT gravity \cite{Chapman:2021eyy}.
 
In this paper, we focus on holographic complexity for two-dimensional dS spacetime in JT gravity. 
As far as we are aware, complexity computations in JT dS$_2$ have been done only using the CV conjecture. One surprising fact in this regard is that holographic volume complexity in dS$_2$ has been found to be upper-bounded by an $\mathcal{O}(1)$ value at early times \cite{Chapman:2021eyy,Jorstad:2022mls}, prior to a sudden infinite jump at a critical time \cite{Jorstad:2022mls}.
However, at first sight, this result might cause confusion from the following perspective. 
Contrary to the late time linear growth of the wormhole volume behind the event horizon of an AdS black hole \cite{Susskind:2014moa,Susskind:2014rva}, the dS space beyond the cosmological horizon of the static patch increases exponentially in time \cite{Susskind:2021esx}. This exponential expansion of dS is one of the reasons why Susskind originally conjectured the growth of holographic complexity in dS to be ``hyperfast'' \cite{Susskind:2021esx}.  In light of this, one might ask why holographic complexity for dS$_2$ does not keep growing as for AdS black holes. 

One simple answer to this question is that in two dimensions there is only one spacelike direction and no ``volume". For instance, in four-dimensional gravity, the black hole horizon is a two-sphere $S^2$, whereas in two dimensions it is just a point. The lack of spatial dimensions prevents the metric from reading the characteristics of the expanding dS space. Does this mean that the features of the expanding universe disappear in two dimensions? No, because there is a dilaton. 

To gain a better grasp on this point, recall that in the original CA papers \cite{Brown:2015bva,Brown:2015lvg} for AdS spacetime in Einstein gravity 
\be 
S =   \frac{1}{16 \pi G} \int  \sqrt{-g} \left(R - 2 \Lambda  \right) ,
\ee
it is pointed out that CA is related to CV as
\be
C \sim \frac{V}{G \ell} \sim \frac{W}{G \ell^2},
\label{CV=CA}
\ee
where $\ell$ is the curvature radius of the background spacetime, $G$ is Newton's constant and $V, W$ is the space and spacetime volume of the WDW patch, respectively. {Schematically, these} are related by  
\be
W \sim \ell \, V \, .
\ee 
In JT gravity, there is a crucial difference due to the dilaton. In this case, {one can never get} the action in the so-called ``Einstein-frame'', rather one always has the ``string-frame'' form of the action: 
\be
S =   \frac{1}{16 \pi G} \int  \sqrt{-g}  \, \phi \left(R - 2 \Lambda  \right) + \cdots  \, ,
\ee
where $\phi$ is the dilaton.
Even after Weyl transformation-like field redefinition 
\be
g_{\mu\nu} \to \Omega(\phi)  
g_{\mu\nu} \, ,
\label{Weylfieldred2}
\ee
one can never get rid of $\phi$ in front of the Ricci scalar in the action. 
Therefore, in JT gravity the relation between action complexity and volume complexity is not given by eq.~\eqref{CV=CA}, but it is rather corrected by the presence of the dilaton as 
\be
C  \sim \frac{\phi_b W}{G \ell^2} , 
\label{CV}
\ee
where $\phi_b$ is the boundary value of the dilaton. 
This is reminiscent of the proposal in \cite{Brown:2018bms}, where the effective Newton's constant is employed for volume complexity in AdS to manifest the expected late-time linear growth. 
The dilaton dependence is more significant in dS spacetime. Indeed, since the dilaton  
diverges at the spacelike future infinity behind the cosmological horizon, it renders action complexity divergent as well. On the other hand, in AdS spacetime the dilaton  
remains finite but the volume grows {linearly} in time.  
The point is that both $\phi_b$ and $W$ matter in JT gravity.

This is consistent with dimensional reduction. In two-dimensional dS, volume complexity is upper bounded at early times, but in higher dimensions, it diverges due to the large expansion of the sphere in the transverse spatial directions. However, after compactification, the growing radius of this transverse spatial sphere reduces to the dilaton in JT gravity. The dilaton is crucial to properly capture the increase in holographic complexity. 

Another subtle issue with volume complexity is that in two-dimensional gravity there is no preferred ``frame'', as we have discussed. We should emphasize that physics must be invariant under the field-redefinition of eq.~\eqref{Weylfieldred2} because fields are not directly observable. Despite the on-shell action being invariant under field-redefinition,  the solution apparently changes. Since in JT gravity, both the dilaton and the metric characterize the solution, it is strange that complexity is defined by the volume, which picks up only the information contained in the metric.  In this regard, it is natural to take action complexity as a starting point to understand the effect of the dilaton in JT gravity. 

The organization of this paper is as follows. In \S \ref{cainjt} we compute action complexity in JT gravity for dS$_2$. For this purpose, an appropriate action is obtained by dimensional reduction starting from Einstein gravity in three dimensions. We show that indeed action complexity correctly captures the effect of the dilaton and gives a result that agrees with what one obtains in three dimensions. In \S \ref{gevwy} we argue that volume complexity  in JT gravity has an ambiguity  by showing that it changes under a field redefinition as in eq.~\eqref{Weylfieldred2}, even though physics should be invariant under this kind of transformation. Further, we show that a specific field redefinition can be found which gives a volume complexity in agreement with the higher-dimensional results.  We end this paper with a discussion in \S \ref{disc}.

\section{CA in JT gravity}
  \label{cainjt}
  
  \subsection{Setup}
  According to  the CA conjecture, the complexity of a state with a holographic realization is proportional to the full gravitational action of the Wheeler-DeWitt (WDW) patch:
  \be
  \mathcal{C}_{\rm A} = \frac{I_{\rm WDW}}{\pi} \, .
  \label{CA}
  \ee 
  We compute action complexity in Jackiw-Teitelboim (JT) gravity,
  whose bulk action is
  \be
  I_{\rm JT} = \frac{1}{8 G} \int_{\mathcal{W}} d^2 x \sqrt{-g} \, \le \phi \, R + L^{-2} \, U \le \phi \ri \ri 
  + \frac{1}{4 G} \int_{\partial \mathcal{W}} dy \sqrt{-h} \, \phi_b \, K \, ,
  \label{JT-action} 
  \ee 
  with $\phi$ the dilaton and $\phi_b$ its value at the boundary $\partial \mathcal{W}$. 
  The dilaton potential is $U(\phi) = 2\phi$ for anti-de Sitter (AdS) spacetime and $U(\phi)=-2 \phi$ for de Sitter (dS) spacetime.
  The boundary term represents the Gibbons-Hawking-York (GHY) action \cite{York:1972sj,Gibbons:1976ue}, where 
  $K = h^{\mu \nu} K_{\mu \nu}$ is the trace of the 
  extrinsic curvature.
  The latter is defined as
  $K_{\mu \nu} = h^{\rho}_{\mu} h^{\s}_{\nu} \nabla_{\rho} n_{\s}$,
  with $n_{\s}$ the outward-directed normal to $\partial \mathcal{W}$.
  
  The equations of motion for the dilaton and the metric are 
  \begin{align}
  	\label{EOM1}
  	R&=-\frac{U'(\phi)}{L^2},\\
  	\label{EOM2}
  	0&=\nabla_a \nabla_b \phi - g_{ab} \nabla^2 \phi +\frac{g_{ab}}{2L^2} U(\phi).
  \end{align}
 In this paper, we work in 
 dS spacetime, so we choose 
  \be
  U(\phi)=-2 \phi \, . 
  \ee
  With this choice, the equations of motion are solved by the following metric and dilaton
  \be
  ds^2 = - f(r) dt^2 + \frac{dr^2}{f(r)} \, , 
  \qquad
  \phi = \frac{r}{L} \, , 
  \label{dS2sol}
  \ee 
  where the blackening factor $f(r)$ is 
  \be
  f(r) = 1 - \frac{r^2}{L^2} \, .
  \label{dsbf}
  \ee
  In the following, 
  we focus on the portion of spacetime with $0 \leq r < \infty$ {(for the reason we shall explain at the beginning of subsection \ref{actd})}.
  Therefore, the position of the cosmological horizon is 
  $r_h = L$.  It is convenient to introduce the tortoise coordinate
  \be
  r^*(r) = \int^r \frac{d\hat{r}}{f(\hat{r})} \, .
  \label{rtdef}
  \ee
  Taking as a boundary condition $r^*(0) = 0$, we get
  \be
  r^*(r) = \frac{L}{2} \log \left| \frac{r + L }{ r - L} \right| \, .\label{rtval}
  \ee
  The Eddington-Finkelstein (EF) coordinates are given by
  \be
  v = t + r^*(r) \, , \qquad
  u = t - r^*(r) \, .
  \label{EF}
  \ee
  To describe the whole of spacetime,
  we need to consider two copies of EF coordinates, 
  one for the right (R) and one for the left (L) side of the Penrose diagram,
  see figure \ref{fig:penrose-dS}.
  Note that $u_R$ is constant along null rays falling into the cosmological horizon, whereas $v_R$ is constant along outgoing null rays.
  The reverse behaviour applies to the EF coordinates for the left side of the Penrose diagram.
  \begin{figure}[h]
  	\centering
  	\begin{tikzpicture}
  		\draw[very thick]  (1.85,1.9) -- (1.85,-1.9) ;
  		\draw[very thick] (-1.9,1.9) -- (-1.9,-1.9);
  		\draw[very thick] (-1.9,1.9) -- (1.85,1.9);
  		\draw[very thick] (-1.9,-1.9) -- (1.85,-1.9);
  		\node[right,rotate=-90] at (2.2, -0.3) {$r=0$};
  		\node[left,rotate=90]  at (-2.25, 1.3) {$r=0$};
  		\draw[dashed, thick, draw= blue!70!white] (-1.9,1.9) to (1.85,-1.9);
  		\draw[dashed, thick, draw= blue!70!white] (-1.9,-1.9) to (1.85,1.9);
  		\node[rotate=-45] at (-0.8,1.2) {$r=L$};
  		\node[rotate=45] at (0.75,1.2) {$r=L$};
  		\node[above] at (0, 1.97) {$r=\infty$};
  		\node[below] at (0, -1.97) {$r=\infty$};
  		\draw[->,very thick, draw=red] (1.85,0)--(1.35,0.5);
  		\node[above left,red] at (1.45,0.4) {$v_R$};
  		\draw[->,very thick, draw=red] (1.85,0)--(2.35,0.5);
  		\node[above right,red] at (2.25,0.4) {$u_R$};
  		\draw[->,very thick, draw=red] (-1.9,0)--(-2.4,-0.5);
  		\node[below left,red] at (-2.3,-0.4) {$u_L$};
  		\draw[->,very thick, draw=red] (-1.9,0)--(-1.5,-0.5);
  		\node[below right,red] at (-1.6,-0.4) {$v_L$};
  		\draw[->, very thick] (3.4,-0.8)--(3.4,0.8);
  		\node[above] at (3.4,0.8) {$t$};
  		\draw[->, very thick] (-3.45,0.8)--(-3.45,-0.8);
  		\node[below] at (-3.45,-0.8) {$t$};
  	\end{tikzpicture}
  	\caption{Penrose diagram for dS spacetime. The time coordinate $t$ runs upwards on the right and downwards on the left. Coordinate axes for the EF coordinates on the right and the left are shown in red.}
  	\label{fig:penrose-dS}
  \end{figure}

  \subsection{Action manual and dimensional reduction}
  \label{actd}

  Action complexity is obtained by evaluating the full gravitational action of the WDW patch as in eq.~\eqref{CA}. The question of which action should be considered is non-trivial since the full JT gravity action of the relevant spacetime region is not just given by $I_{\rm JT}$ in eq.~\eqref{JT-action}. In fact, there are several additional boundary contributions. In order to determine the right action, we exploit the fact that the dS JT gravity action in two dimensions can be obtained by dimensional reduction from three-dimensional pure dS, starting from the Einstein-Hilbert (EH) action plus boundary terms \cite{Svesko:2022txo}.\footnote{The dimensional reduction allows us to circumvent the study of the variational problem in JT gravity, leading to the same result for the full gravitational action.}   
  We employ this procedure to explicitly compute action complexity in dS$_2$. 
  
 Before we proceed, we point out that there are two ways to obtain dS$_2$ JT gravity from dimensional reduction. One is half reduction from pure dS$_3$ \cite{Svesko:2022txo} and the other is full reduction from Schwarzschild-dS$_d$ ($d \ge 4$) in the near-Nariai limit \cite{Nariai1950, Ginsparg:1982rs} (see \cite{Anninos:2012qw, Maldacena:2019cbz, Svesko:2022txo} as well). For our analysis, we use half reduction from pure dS$_3$, which automatically reduces to the $r \ge 0$ range. The dS$_2$ spacetime obtained from Schwarzschild-dS$_d$ ($d \ge 4$) in the near-Nariai limit also includes $r < 0$ and a horizon at negative $r$. In this limit, we are zooming in on the very tiny region near the black hole and cosmological horizons, and the $r$-negative horizon corresponds to the original black hole one.  Since our main interest is holographic complexity beyond the cosmological horizon, rather than black hole complexity, we focus on $r > 0$. For this purpose, half reduction from three-dimensional pure dS is enough.  
  In three dimensions, the full gravitational action is \cite{Lehner:2016vdi}
  \be
  \begin{aligned}
  	I_{\rm tot} &=   I_{\rm EHGHY}  +   I_{\rm null}  +I_{\rm joint} +  I_{\rm ct}  \, , \\
  	  I_{\rm EHGHY}  &\equiv \frac{1}{16 \pi G_{(3)}} \int_{\tilde{\mathcal{W}}} d^3 X \sqrt{-g_{(3)}} \le R_{(3)} - 2 \Lambda_{(3)} \ri
  	+ \frac{1}{8 \pi G_{(3)}} \int_{\partial \tilde{\mathcal{W}}} d^2 Y \sqrt{-h_{(3)}} \, K_{(3)} \, , \\
  	  I_{\rm null} &\equiv \frac{1}{8 \pi G_{(3)}} \int_{\tilde{\mathcal{B}}} dS \, d\l \, \sqrt{\gamma} \, k \, , \qquad 
  	I_{\rm joint} \equiv \frac{1}{8 \pi G_{(3)}} \int_{\tilde{\mathcal{J}}} d\theta \, \sqrt{\gamma} \, \tilde{\mathfrak{a}}  \, , \\
  	  I_{\rm ct} &\equiv \frac{1}{8 \pi G_{(3)}} \int_{\tilde{\mathcal{B}}} dS \, d\l \, \sqrt{\gamma} \, \Theta \, \log \left| \tilde{L} \, \Theta \right| \, .
  \end{aligned}
  \label{starting-action}
  \ee
 The term $  I_{\rm EHGHY} $ contains the EH and the GHY actions, the latter coming from spacelike and timelike codimension-one boundaries. 
  Contributions from codimension-one null surfaces $\tilde{\mathcal{B}}$ are given by $ I_{\rm null}$  \cite{Lehner:2016vdi,Parattu:2015gga}. 
  In $ I_{\rm joint}$, we have the contributions from codimension-two joints $\tilde{\mathcal{J}}$ \cite{Hayward:1993my,Lehner:2016vdi} at the intersection between codimension-one surfaces.  
  Finally, $ I_{\rm ct}$ is a counterterm for null boundaries $\tilde{\mathcal{B}}$ \cite{Lehner:2016vdi} which ensures invariance of the full action under reparameterization of null normals.
  In what follows, we analyze each term separately
  and we apply dimensional reduction to determine the right action for holographic complexity in dS$_2$ JT gravity.

  The EH-GHY action in three dimensions with positive cosmological constant reads
  \be
  \begin{aligned}
  I_{\rm EHGHY} &= \frac{1}{16 \pi G_{(3)}} \int_{\tilde{\mathcal{W}}} d^3 X \sqrt{-g_{(3)}} \le R_{(3)} - 2 \Lambda_{(3)} \ri \\
  &+ \frac{1}{8 \pi G_{(3)}} \int_{\partial \tilde{\mathcal{W}}} d^2 Y \sqrt{-h_{(3)}} \, K_{(3)} \, ,
  \end{aligned}
  \label{EH3}
  \ee
   where 
    \be
  \Lambda_{(3)} = \frac{1}{L_{(3)}^2} \,.
  \ee
 This admits a three-dimensional dS spacetime solution
  \begin{align}
  ds_{(3)}^2 = g_{(3) M N} dX^{M} dX^{N} = - f(r) dt^2 + \frac{dr^2}{f(r)} + r^2 d\theta^2 \,, \quad
  f(r)  = 1 - \frac{r^2}{L_{(3)}^2} \,.
  \label{dS3sol}
  \end{align}
 In the GHY boundary term in eq.~\eqref{EH3}, $h_{(3)}$ is the determinant of the induced metric on $\partial \tilde{\mathcal{W}}$ and $K_{(3)}$ denotes the trace of the extrinsic curvature. \\
  We now introduce the metric ansatz 
  \be
  ds_{(3)}^2 = g_{(2) \mu \nu} dx^{\mu} dx^{\nu} + L_{(3)}^2 \phi^2(x) d\theta^2 \, ,
  \label{metric-ansatz}
  \ee
  where we have expressed $X^{M} = \le t, r, \theta \ri = \le x^{\mu}, \theta \ri$ with $M=0,1,2$ and $\mu = 0,1$. As in eq.~\eqref{dS2sol}, we consider a solution for the dilaton $\phi(x)$ depending just on the radial coordinate: $\phi = \phi(r)$.
  From the metric ansatz \eqref{metric-ansatz}, we get \cite{Svesko:2022txo}
  \be
  \begin{aligned}
  	R_{(3)} &= R_{(2)} - \frac{2}{\phi} \square_{(2)} \phi \, , \\
  	K_{(3)} &= K_{(2)} + \frac{1}{\phi} n^{\mu} \nabla_{(2) \mu} \phi \, , \\
  	\sqrt{-g_{(3)}} &= L_{(3)} \, \phi \, \sqrt{- g_{(2)}} \, ,
  \end{aligned}
  \ee
  with $n^{\mu}$ the normal vector to the two-dimensional boundary $\partial \tilde{\mathcal{W}}$. 
  Plugging into eq.~\eqref{EH3}, we obtain
  \be
  \begin{aligned}
  	I_{\rm JT} &= \frac{L_{(3)}}{8 G_{(3)}} \int_{\mathcal{W}} d^2 x \sqrt{-g_{(2)}} \left[ \phi \le R_{(2)} - 2 \Lambda_{(3)} \ri - 2 \, \square_{(2)} \phi \right] \\
  	&+ \frac{L_{(3)}}{4 G_{(3)}} \int_{\partial \mathcal{W}} dy \sqrt{-h_{(2)}} \, \le \phi \, K_{(2)} + n^{\mu} \nabla_{(2) \mu} \phi \ri \, ,
  \end{aligned}
  \label{JT-red}
  \ee
  where $\mathcal{W}$ and $\partial \mathcal{W}$ are the manifolds endowed with metric $g_{(2) \mu \nu}$ and $h_{(2) \mu \nu}$, respectively.
 Defining $G \equiv G_{(2)} = G_{(3)}/L_{(3)}$ and $L \equiv L_{(3)}$, we recognize the JT gravity action $I_{\rm JT}$ in eq.~\eqref{JT-action} with some additional terms.  Three comments regarding the EH-GHY contributions are in order:  
\begin{itemize}
\item  According to the CA conjecture, we must evaluate the full gravitational action of the WDW patch. Such a spacetime region, which we denote by $\mathcal{W}$, is bounded by null codimension-one surfaces.
  Since there are neither timelike nor spacelike boundaries, the GHY term in eq.~\eqref{JT-red} vanishes.
  \item With the metric ansatz in eq.~\eqref{metric-ansatz}, the three-dimensional dS spacetime solution in eq.~\eqref{dS3sol} comes down to the two-dimensional dS solution with linear dilaton given in eqs.~\eqref{dS2sol} and \eqref{dsbf}.  
  On-shell we have 
  \be
  R_{(2)} = 2 \Lambda_{(3)} = \frac{2}{L_{(3)}^2} \, ,
  \ee
  so the bulk contribution $I_{\rm JT}$ in eq.~\eqref{JT-red} reduces to  
  \be
  I_{\rm bulk} = 
   - \frac{1}{4 G_{(2)}} \int_{\mathcal{W}} d^2 x \sqrt{-g_{(2)}} \, \square_{(2)} \phi \, .
  \label{bulk_action}
  \ee
  Even though we refer to this expression as two-dimensional ``bulk'' action, $I_{\rm bulk}$ is in fact a boundary term since it comprises a total divergence.\footnote{As it will be clear later, at late times all contributions to the full action have the same divergence structure. Therefore, by neglecting $I_{\rm bulk}$ the qualitative behavior of action complexity at late times does not change.} 
   \item We can integrate the term $\square_{(2)} \phi$ getting a codimension-one contribution for the null boundaries of the WDW patch distinct from $I_{\rm null}$ in eq.~\eqref{starting-action}:
 \be
 I_{\rm bulk} = - \frac{1}{4 G_{(2)}} \int_{\tilde{\mathcal{B}}} d\l \, k^{\mu} \nabla_{(2) \mu} \phi = 
 - \frac{1}{4 G_{(2)}} \int_{\tilde{\mathcal{B}}} d\l \, \partial_{\l} \phi \, ,
 \label{bulk-boundary}
 \ee
 where $k^{\mu}$ is the null normal to $\tilde{\mathcal{B}}$,
 see \cite{Goto:2018iay} for an analog analysis in AdS$_2$ JT gravity. 
 As we will see, this term yields nonzero contribution and is important to get consistency with the three-dimensional CA result.
        \end{itemize}
  Besides the bulk term,
  the full gravitational action of the WDW patch contains boundary terms for null codimension-one surfaces and joint terms for codimension-two surfaces at the intersection of null bounding surfaces, which we consider next.

    The contribution from null boundaries
  has been originally studied in 
  \cite{Lehner:2016vdi,Parattu:2015gga}.
  In three-dimensional spacetime it is 
  \be
  I_{\rm null} = \frac{1}{8 \pi G_{(3)}} \int_{\tilde{\mathcal{B}}} dS \, d\l \, \sqrt{\gamma} \, k \, , 
  \label{null-bou}
  \ee 
  where $\l$ is a parameter running along the null geodesics generating the surface $\tilde{\mathcal{B}}$, 
  $S$ is the transverse spatial direction to such generators,
  and $\gamma$ is the determinant of the induced metric in the $S$ direction.
  The constant $\kappa$ is defined by the geodesic equation
  \be
  \tilde{k}^M \nabla_{(3) M} \tilde{k}^N =\kappa \,  \tilde{k}^N \, ,
  \ee
  with $\tilde{k}^M = \frac{dX^M (\l)}{d\l}$ the null generator.
  In other words,
  $\kappa$ measures the failure of $\l$ to be an affine parameter.
  Consequently, 
  we can set $\kappa=0$ by a wise parameterization choice,
  getting rid of the $I_{\rm null}$ contribution.

  At the intersection between bounding codimension-one surfaces,
  where the boundary is non-smooth,
  the joint term comes into play.
  The joint contributions involving just timelike and spacelike intersecting surfaces have been investigated in \cite{Hayward:1993my},
  while joints involving at least one null boundary have been studied in \cite{Lehner:2016vdi}.
  In our computation
  we will meet only joints involving two null boundaries, which
  in three-dimensional spacetime are given by
  \be
  \begin{aligned}
  I_{\rm joint} &= \frac{1}{8 \pi G_{(3)}} \int_{\tilde{\mathcal{J}}} d\theta \, \sqrt{\gamma} \, \tilde{\mathfrak{a}} \, , \\
   \sqrt{\gamma} = \sqrt{g_{(3) \theta \theta}}&  \, , 
   \qquad
  \tilde{\mathfrak{a}} = {\rm sign}({\rm joint}) \times \log \left| \frac{\mathbf{\tilde{k}_1} \cdot \mathbf{\tilde{k}_2}}{2} \right|_{\tilde{\mathcal{J}}} \, ,
  \label{joint-start}
  \end{aligned}
  \ee
  where $\mathbf{\tilde{k}_1}, \mathbf{\tilde{k}_2}$ denote the one-forms normal to null boundaries, which are taken to be outward-directed from the spacetime region of interest.
 The sign of the joint term is fixed by the following rule. Choosing either of the two null boundaries intersecting at the joint, the sign of the corresponding contribution \eqref{joint-start} is positive if the bulk region $\mathcal{W}$ is at the future (past) of the segment and the joint itself is located at the past (future) edge of the selected segment. In the remaining configurations, the joint term has a negative sign \cite{Lehner:2016vdi}.

    Applying the dimensional reduction, we obtain
  \be
  I_{\rm joint} = \frac{1}{4 G_{(2)}} \sum_{\mathcal{J}} \phi(r_\mathcal{J}) \,  \mathfrak{a}_{\mathcal{J}} \, ,
  \qquad
  \mathfrak{a}_{\mathcal{J}} = {\rm sign}({\rm joint}) \times \log \left| \frac{\mathbf{k_1} \cdot \mathbf{k_2}}{2} \right|_{J} \, .
  \label{joint_term}
  \ee
  Here $r = r_\mathcal{J}$ is the radial position of the joint,
  and we consider one-forms $\mathbf{k_1}$, $\mathbf{k_2}$ normal to one-dimensional boundaries.
  By spherical symmetry of the dS$_3$ geometry, we trivially have $\mathbf{\tilde{k}_i} = \mathbf{k_i}$ and $\tilde{k}_i^{M} = \le k_i^{\mu}, 0 \ri$, with $i=1,2$.

  The joint contributions in eq. (\ref{joint_term}) are affected by the arbitrariness of choosing the normalization of null normals $\mathbf{k_1}, \mathbf{k_2}$. Such ambiguity can be partially removed by requiring that $\mathbf{k_i} \cdot \partial_t = \pm \a$ at the spacetime boundary \cite{Carmi:2016wjl,Chapman:2016hwi,Carmi:2017jqz}, with $\partial_t$ the timelike Killing vector in the boundary theory and $\a$ a positive constant.
  Still, the constant $\a$ can be arbitrarily chosen. For the gravitational action to be invariant under the reparameterization of null generators, a counterterm must be added for each null boundary \cite{Lehner:2016vdi}.
  In three-dimensional spacetime, the counterterm has the following form
  \be
  I_{\rm ct} = \frac{1}{8 \pi G_{(3)}} \int_{\tilde{\mathcal{B}}} dS \, d\lambda \, \sqrt{\gamma} \, \Theta \, \log \left| \tilde{L} \, \Theta \right| \, ,
  \ee
  where $\tilde{L}$ is an arbitrary length scale.
  In the above expression, $\Theta$ is the expansion of null geodesics given by $\Theta = \partial_\l \log \sqrt{\gamma}$.
  An explicit calculation with the metric ansatz (\ref{metric-ansatz}) leads to
  \be
  \Theta = \partial_\l \log \sqrt{g_{(3) \theta \theta}} 
  = \partial_\l \log \phi 
  = \frac{\partial_\l \phi}{\phi} \, .
  \ee
  Therefore,
  in JT gravity the counterterm reads 
  \be
  I_{\rm ct} = \frac{1}{4 G_{(2)}} \int_\mathcal{B} d\lambda \, \partial_\l \phi \, \log \left| \tilde{L} \, \partial_\l \log \phi \right|  \, .
  \label{counterterm}
  \ee

  Summarizing, the full gravitational action for holographic complexity in dS$_2$ JT is given by
  \be
  \begin{aligned}
  	I_{\rm JT dS_2} &=  I_{\rm bulk} + I_{\rm joint} + I_{\rm ct} \\
	&= - \frac{1}{4 G_{(2)}} \int_{\tilde{\mathcal{B}}} d\l \, \partial_{\l} \phi 
  	+ \frac{1}{4 G_{(2)}} \sum_{\mathcal{J}} \phi(r_\mathcal{J}) \, \mathfrak{a}_\mathcal{J}  \\
  	&\, \, \, \, \, \, \, + \frac{1}{4 G_{(2)}} \int_\mathcal{B} d\lambda \, \partial_\l \phi \, \log \left| \tilde{L} \, \partial_\l \log \phi \right|
  	\, .
  \end{aligned}
  \label{tot_action}
  \ee

The lesson we learn from the dimensional reduction analysis is that the effects of the dilaton, which explicitly appears in all terms of the full action \eqref{tot_action}, are important for the evaluation of holographic complexity. This is consistent with the discussion in the introduction. Note that in JT gravity the value of dilaton varies, and as a result, one has an effective Newton ``constant"
\be
\frac{1}{G_{ (2) \, {\rm eff} }} \equiv \frac{\phi}{G_{(2)} } \, .
\ee
  We recall that in two dimensions there is no ``area'' for a black hole horizon, but instead, there is a dilaton. The dilaton plays the role of area. In fact, in $d  (\ge 3)$-dimensional dS spacetime, the orthogonal $S^{d-2}$ area grows to infinity near the future spacelike infinity, and by dimensional reduction the size factor of the $S^{d-2}$ area becomes the dilaton. Since the divergence of complexity in higher-dimensional dS is associated with the growth of the orthogonal $S^{d-2}$ area \cite{Jorstad:2022mls}, without taking into account the dilaton one cannot see the divergence of complexity in two-dimensional dS spacetime in JT gravity.

  \subsection{Action evaluation}

  We now move to the explicit computation of the full gravitational action,
  mainly following \cite{Jorstad:2022mls}.
  To ease the notation, we set $L_{(3)} \equiv L$.
  First, 
  we introduce the stretched horizons for both the left and the right
  sides of the Penrose diagram. These are $r$-constant surfaces described by
  \be
  r_{st} = \rho L \, , \qquad
  0 < \rho < 1 \, ,
  \label{stretched}
  \ee
  where the limit $\rho \sim 1$ is intended. 
  We then attach the WDW patch to the two stretched horizons,
  and we define the anchoring times as
  $t_L$ and $t_R$ for the left and the right horizons, respectively. 
  It is not restrictive to focus on the symmetric case $t_R = - t_L$.
  For convenience, we define a dimensionless time as $t_R = L \, \tau$. 
  We focus on the case where 
	the tip of the WDW patch does not meet the future spacelike infinity at $r=\infty$.
  \\
  \begin{figure}[h]
  	\centering
  	\begin{tikzpicture}
  		\draw[very thick]  (1.9,1.9) -- (1.9,-1.9) ;
  		\draw[very thick] (-1.9,1.9) -- (-1.9,-1.9);
  		\draw[very thick] (-1.9,1.9) -- (1.9,1.9);
  		\draw[very thick] (-1.9,-1.9) -- (1.9,-1.9);
  		\node[right,rotate=-90] at (2.2,0.5) {$r=0$};
  		\node[left,rotate=90]  at (-2.25,0.5) {$r=0$};
  		\draw[draw=red,  thick] (1.9,1.9)..controls (0.8,0.5) and (0.8,-0.5) .. (1.9,-1.9);
  		\draw[draw=red,  thick] (-1.9,1.9)..controls (-0.8,0.5) and (-0.8,-0.5) .. (-1.9,-1.9);
  		\draw[very thick,red,fill=white!70!red] (-1.1,0.3)--(0,1.4)--(1.1,0.3)--(0,-0.8)--(-1.1,0.3);
  		\draw[dashed] (0,1.9)--(0,-1.9);
  		\draw[dashed, thick, draw= blue!70!white] (-1.9,1.9) to (1.9,-1.9);
  		\draw[dashed, thick, draw= blue!70!white] (-1.9,-1.9) to (1.9,1.9);
  		\node[above] at (0, 1.97) {$r=\infty$};
  		\node[below] at (0, -1.97) {$r=\infty$};
  		\node[above] at (0,1.4) {$r_+$};
  		\node[below] at (0,-0.8) {$r_-$};
  		\node[right] at (1.1,0.3) {$\tau$};
  		\node[left] at (-1.1,0.3) {$\tau$};
  		\node[above left,rotate=45,red] at (-0.3,1.2) {FL};
  		\node[above right,rotate=-45,red] at (0.3,1.2) {FR};
  		\node[below left,rotate=-45,red] at (-0.3,-0.5) {PL};
  		\node[below right,rotate=45,red] at (0.3,-0.5) {PR};
  		\node at (0.3,0.7) {III};
  		\node at (0.5,0.1) {II};
  		\node at (0.15,-0.4) {I};
  	\end{tikzpicture}
  	\caption{Penrose diagram for dS spacetime showing the WDW patch (red region). The stretched horizons at $r = \rho L$ are represented by red curves.}
  	\label{fig:WDW-dS}
  \end{figure}

  For the right side of the geometry, 
  the future boundary of the WDW patch is at constant $u_R$,
  whereas the past boundary is at constant $v_R$.
  From eq.~\eqref{EF}, the defining equations are
  \be
  \begin{aligned}
  	u_{FR} &= t_R - r^*(r_{st}) = L \, \tau - \frac{L}{2} \log \le \frac{1+ \rho }{1- \rho} \ri \, ,  \\
  	v_{PR} &= t_R + r^*(r_{st}) = L \, \tau + \frac{L}{2} \log \le \frac{1+ \rho}{ 1- \rho} \ri \, .
  \end{aligned}
  \label{WDW-boundaries}
  \ee
  Let us denote by $r=r_{\pm}$ the position of the future and past tip of the WDW patch, respectively.
  By symmetry, both tips are located on the vertical axis in the middle
  of the Penrose diagram, which is described by $t=0$.
  So, evaluating $u_{FR}$ defined in eq.~\eqref{WDW-boundaries} at the future tip, from eq.~\eqref{EF} we get
  \be
  L \, \tau - \frac{L}{2} \log \le \frac{1+ \rho }{1- \rho} \ri 
  = - \frac{L}{2} \log \le \frac{r_+ + L}{r_+ - L} \ri \, ,
  \label{for-r+}
  \ee
  which leads to
  \be
  \frac{r_+}{L} = \frac{\cosh \tau - \rho \sinh \tau}{\rho \cosh \tau - \sinh \tau} \, .
  \label{r+}
  \ee
  Similarly, computing $v_{PR}$ defined in eq.~\eqref{WDW-boundaries} at the past tip we obtain
  \be
  L \, \tau + \frac{L}{2} \log \le \frac{1+ \rho}{ 1- \rho} \ri
  = \frac{L}{2} \log \frac{r_- + L}{r_- - L} \, ,
  \label{for-r-}
  \ee
  from which
  \be
  \frac{r_-}{L} = \frac{\cosh \tau + \rho \sinh \tau}{\rho \cosh \tau + \sinh \tau} \, .
  \label{r-}
  \ee
  The time $\tau = \tau_{\infty}$ at which the future tip meets the future spacelike infinity $r \to \infty$ can be obtained by eq.~\eqref{r+} as
  \be
  \tau_{\infty} = \tanh^{-1} \rho \, .
  \label{tau-infty}
  \ee
  We can thus rewrite eqs. (\ref{for-r+}) and (\ref{for-r-}) as
  \be
  \tau_{\infty} - \tau
  = \tanh^{-1} \le \frac{L}{r_+} \ri \, , \qquad
  \tau + \tau_{\infty}
  = \tanh^{-1} \le \frac{L}{r_-} \ri \, ,
  \label{tau-rpm}
  \ee
  or equivalently
  \be
  r_{\pm} = L \coth \le \tau_{\infty} \mp \tau \ri \, .
  \label{rpm-tau}
  \ee
  Since we are interested in how complexity grows as $\tau$ approaches $\tau_{\infty}$, we focus on $\tau \leq \tau_{\infty}$.
  We now evaluate the gravitational action in eq.~\eqref{tot_action} term by term.

  Explicit computation of the boundary term in eq.~\eqref{bulk-boundary} gives
  \be
  \begin{aligned}
  I_{\rm bulk} &=  \left. \frac{\phi}{2 G_{(2)}} \right|^{r_+}_{\rho L} + \left. \frac{\phi}{2 G_{(2)}} \right|^{r_-}_{\rho L} = \frac{1}{G_{(2)}} \le \frac{r_+ + r_- - 2 \rho L}{2 L}  \ri \\
  &= \frac{1}{G_{(2)}} \, \frac{\rho \le 1 - \rho^2 \ri \cosh^2 \tau }{1- \le 1- \rho^2 \ri \cosh^2 \tau} \, ,
  \end{aligned}
  \label{bulk-final}
  \ee
  where in the last step eqs.~\eqref{r+} and \eqref{r-} has been used.

  We now move to the boundary terms. 
  The null normals $\mathbf{k} = k_{\mu} dx^{\mu}$ to the four codimension-one boundaries of the WDW patch are
  \be
  \begin{aligned}
  	\mathbf{k}_{FR} &= \a \, du_R = \a \le dt - \frac{dr}{f(r)} \ri \, , \\
  	\mathbf{k}_{PR} &= - \b \, dv_R = \b \le - dt - \frac{dr}{f(r)} \ri \, , \\
  	\mathbf{k}_{FL} &= -\a' \, dv_L = \a' \le - dt - \frac{dr}{f(r)} \ri \, , \\
  	\mathbf{k}_{PL} &= \b' \, du_L = \b' \le dt - \frac{dr}{f(r)} \ri \, , 
  \end{aligned}
  \ee
  where $\a, \b, \a', \b'$ are positive arbitrary constants
  and F (P) stands for future (past),
  see figure \ref{fig:WDW-dS}.
  This choice corresponds to an affine parameterization  of null normals, 
  thus the boundary term \eqref{null-bou} for null surfaces vanishes.
  To compute the four joint terms in eq.~\eqref{joint_term} we need
  \be
  \begin{aligned}
  	\mathfrak{a}_{+} &= + \log \left| \frac{\mathbf{k}_{FR} \cdot \mathbf{k}_{FL}}{2} \right|_{r = r_+}
  	= \log  \left| \frac{\a \a'}{f(r_+)} \right| \, , \\
  	\mathfrak{a}_{-} &= + \log \left| \frac{\mathbf{k}_{PR} \cdot \mathbf{k}_{PL}}{2} \right|_{r = r_-}
  	= \log  \left| \frac{\b \b'}{f(r_-)} \right| \, , \\
  	\mathfrak{a}_{st,R} &= - \log \left| \frac{\mathbf{k}_{FR} \cdot \mathbf{k}_{PR}}{2} \right|_{r = r_{st}}
  	= - \log  \left| \frac{\a \b}{f(\rho L)} \right| \, , \\
  	\mathfrak{a}_{st,L} &= - \log \left| \frac{\mathbf{k}_{FL} \cdot \mathbf{k}_{PL}}{2} \right|_{r = r_{st}}
  	= - \log  \left| \frac{\a' \b'}{f(\rho L)} \right| \, .
  \end{aligned}
  \ee
  Then, the total joint contribution is
  \be
  \begin{aligned}
  	I_{\rm joint} &= I_+ + I_- + I_{st,R} + I_{st,L} \\
  	&= \frac{1}{4 G_{(2)} L} \le r_+ \, \log  \frac{\a \a' L^2}{r_+^2 - L^2} +  r_- \, \log  \frac{\b \b' L^2}{r_-^2 - L^2} - \rho L \, \log \frac{\a \b \a' \b'}{\le 1- \rho^2 \ri^2} \ri \, .
  \end{aligned}
  \label{joint-final}
  \ee
  
  Finally, we have to calculate the counterterm for null boundaries given by eq. (\ref{counterterm}).
  We start from the FR boundary. The null vector orthogonal to this surface is
  \be
  k_{FR}^{\mu} = g^{\mu \nu} k_{FR \, \nu} 
  = - \a \le \frac{1}{f(r)} \, , 1 \ri \, .  
  \ee
  By definition, the null vector is  $k^{\mu} = \partial_{\l} x^{\mu} = \le \dot{t}, \dot{r} \ri$,
  where the dot denotes a derivative with respect to the affine parameter $\l$.
  We thus conclude that $\dot{r} = - \a $, implying that $r = -\alpha \, \l$ is an affine parameter too which increases when $\l$ decreases. 
  Putting all together, we get
  \be
  \begin{aligned}
  	I_{ct, FR} &= 
  	\frac{1}{4 G_{(2)} L} \int_{r_+}^{\rho L} dr \, \log \frac{\tilde{L} \, \a}{r} \\
  	&= \frac{1}{4 G_{(2)} L} \le \rho L - r_+ + \rho L \log \le \frac{\tilde{L} \, \a}{\rho L} \ri - r_+ \log \le \frac{\tilde{L} \, \a}{r_+} \ri \ri \, .
  \end{aligned}
  \ee
  By symmetry, the contribution on the future left boundary is simply 
  $I_{ct, FL} = I_{ct, FR} \le \a \rightarrow \a' \ri$.
  Similarly, the counterterm for the right past boundary is
  \be
  \begin{aligned}
  	I_{ct, PR}  &= 
  	 \frac{1}{4 G_{(2)} L} \int_{r_-}^{\rho L} dr \, \log \frac{\tilde{L} \, \b}{r} \\
  	&= \frac{1}{4 G_{(2)} L} \le \rho L - r_- + \rho L \log \le \frac{\tilde{L} \, \b}{\rho L} \ri - r_- \log \le \frac{\tilde{L} \, \b}{r_-} \ri \ri \, ,
  \end{aligned}
  \ee
  and the contribution from the past left boundary is $I_{ct, PL} = I_{ct, PR} \le \b \rightarrow \b' \ri$. 
  So, the total counterterm reads
  {
  \footnotesize
  \be
  \begin{aligned}
  	I_{ct} &= I_{ct, FR} + I_{ct, FL} + I_{ct, PR} + I_{ct, PL} \\
  	&= \frac{1}{4 G_{(2)} L} \le 4 \rho L - 2 r_+ - 2 r_- + \rho L \log \le \frac{\tilde{L}^4 \a \b \a' \b'}{\rho^4 L^4} \ri - r_+ \log \le \frac{\tilde{L}^2 \a \a'}{r_+^2} \ri - r_- \log \le \frac{\tilde{L}^2 \b \b'}{r_-^2} \ri \ri \, .
  \end{aligned}
  \label{ct-final}
  \ee
  }
  Note that the terms containing the arbitrary constants $\a, \b, \a', \b'$ cancel the corresponding terms in the joint contribution (\ref{joint-final}).
  In details,
  {
  	\footnotesize
  	\be
  	\begin{aligned}
  		I_{\rm joint} + I_{ct} &= \frac{1}{2 G_{(2)}L} \le 2 \rho L - r_+ - r_- + r_+ \, \log  \frac{r_+ L}{\tilde{L} \sqrt{ r_+^2 - L^2 }} +  r_- \, \log  \frac{r_- L}{\tilde{L} \sqrt{ r_-^2 - L^2 }} - \rho L \, \log \frac{\rho^2 L^2}{\tilde{L}^2 \le 1- \rho^2 \ri} \ri \\
  		&= \frac{1}{2 G_{(2)}L} \le L \coth \le \tau_{\infty} - \tau \ri \le \log \le \frac{L \cosh \le \tau_{\infty} - \tau \ri }{\tilde{L}} \ri -1 \ri + \le \tau \rightarrow - \tau \ri \right. \\
  		&\left. - 2 L \tanh \le \tau_{\infty} \ri \le \log \le \frac{L \sinh \le \tau_{\infty} \ri }{\tilde{L}} \ri -1 \ri \ri \, ,
  	\end{aligned}
  	\label{joint+ct}
  	\ee
  }
  where in the second equality eqs.~\eqref{tau-infty} and \eqref{rpm-tau} have been used.

  Summing up eqs.~\eqref{bulk-final} and \eqref{joint+ct}, we finally get the action complexity
  \be
  \begin{aligned}
  	\mathcal{C}_{\rm A} &= \frac{I_{\rm JTdS_2}}{\pi} = 
  	\frac{I_{\rm bulk} + I_{\rm joint} + I_{ct}}{\pi} \\
  	&= \frac{1}{2 \pi G_{(2)}} \le \frac{2 \rho \le 1 - \rho^2 \ri \cosh^2 \tau }{1- \le 1- \rho^2 \ri \cosh^2 \tau} - 2  \tanh \le \tau_{\infty} \ri \le \log \le \frac{L \sinh \le \tau_{\infty} \ri }{\tilde{L}} \ri -1 \ri \right. \\
  	&\left. + \coth \le \tau_{\infty} - \tau \ri \le \log \le \frac{L \cosh \le \tau_{\infty} - \tau \ri }{\tilde{L}} \ri -1 \ri + \le \tau \rightarrow - \tau \ri \ri \, ,
	\label{finalCA}
  \end{aligned}
  \ee
  which matches the dS$_3$ result found in \cite{Jorstad:2022mls},
  with $G_{(2)} = G_{(3)}/L$.

Some comments are in order:
\begin{itemize}
\item  The $\mathcal{C}_{\rm A} $ given by eq.~\eqref{finalCA} diverges at the critical time
  \be
  \cosh^2 \tau_{\infty} = \frac{1}{1- \rho^2} \, ,
  \ee
  which is exactly eq.~\eqref{tau-infty}. In the limit $\tau \to \tau_{\infty}$, both $r_+$ and $\mathcal{C}_{\rm A}$ diverge at the leading order. In particular, 
    \be
  r_+  =  \frac{L}{\left( \tau_\infty - \tau \right)} +  \mbox{(subleading)} \qquad \mbox{as $\tau \to \tau_\infty$} \, ,
  \ee
  where we have used eq.~\eqref{r+}, and 
    \be
  \mathcal{C}_{\rm A} = \frac{1}{\pi G_{(2)}} \frac{1}{\sqrt{2 - \rho^2}} \frac{1}{\left( \tau - \tau_\infty \right)} \propto \frac{r_+}{G_{(2)}} +  \mbox{(subleading)} \qquad \mbox{as $\tau \to \tau_\infty$} \, .
  \ee
Therefore, the late-time behavior of $\mathcal{C}_{\rm A}$ is 
\be
\mathcal{C}_{\rm A} \sim \frac{\phi}{G_{(2)}} \equiv \frac{1}{G_{ (2) \, {\rm eff} }} \, .
\ee
As discussed in the introduction, this is exactly what we expect in JT gravity. 
\item As first pointed out in \cite{Susskind:2021esx}, in $d \ge 3$ dimensions volume complexity diverges at the critical time. This can be understood from the fact that maximal slices anchored at the two stretched horizons bend upwards to the future spacelike infinity, where local $(d-2)$-spheres exponentially expand. In JT gravity, the effect of such an exponential expansion appears in the linear dilaton. Note that the static coordinate $r$ behaves as time beyond the cosmological horizon $r \ge L$ and it diverges at the future spacelike infinity. 
\item Even though $\tau_\infty$ diverges in the limit $\rho \to 1$, it does as 
\be
\rho \equiv 1 - \epsilon \,, \quad  \tau_\infty =  \frac{1}{2} \log \frac{2}{\epsilon} + O(\epsilon) \,.
\ee 
So, the $\tau_\infty$ dependence on the stretched horizon parameter $\epsilon$ is very mild  \cite{Susskind:2021esx,Jorstad:2022mls}. 
\item Without taking into account the dilaton, volume complexity in dS$_2$ remains finite due to the lack of a local exponentially expanding $(d-2)$-sphere in two dimensions \cite{Chapman:2021eyy,Jorstad:2022mls}.
\end{itemize}

\section{Geodesics, Volume and Weyl dependence}
\label{gevwy}

\subsection{Weyl field redefinition}

In this section, we discuss volume complexity and its subtleties. Before we proceed, we review the basic point.

The most general two-dimensional dilaton gravity bulk action up to two derivatives can be written in the form
\begin{align}
	S=\frac{1}{8 G}\int_{\mathcal{W}}  d^2 x \sqrt{-\tilde{g}}\left(U_1(\Phi)\tilde{R}+U_2(\Phi)\tilde{g}^{\mu\nu}\tilde{\nabla}_\mu\Phi\tilde{\nabla}_\nu\Phi+ { {L^{-2}}} U_3(\Phi)\right) \, .
\end{align}
We can perform a Weyl field redefinition
\be
\tilde{g}_{\mu \nu} = e^{2 \omega} \, g_{\mu \nu} \, , 
\qquad
\nabla_{\mu} \omega = - \frac{U_2(\Phi)}{2 U'_1(\Phi)} \nabla_{\mu} \Phi \, ,
\ee
to get rid of the kinetic term.
Then, by doing a simple field redefinition $\phi=U_1(\Phi)$, we can remove $U_1$ as well. 
Note that this Weyl field redefinition does not change the coefficient of the $\sqrt{-\tilde{g}}\tilde{R}$ terms and we are left with just one function of $\phi$ which is related to $U_3$ as $U(\phi)=U_3(\Phi)$. Therefore, a general dilation gravity system can be brought to the form
\begin{align}
	S=\frac{1}{8 G}\int d^2x \sqrt{-g}  \left(\phi R+ L^{-2} U(\phi) \right) \, ,
\end{align}
which admits the two-dimensional dS solution with linear dilaton in eq.~\eqref{dS2sol}. 

Note that even though the intermediate Weyl transformation done above is {\it just a field redefinition}, and hence should not affect physical quantities, it has significant influence in the context of the CV conjecture, since the volume changes by Weyl transformation. 
In this section, we study the effects of Weyl transformations on the CV conjecture.  
To evaluate volume complexity, it is convenient  to use the metric ansatz 
\be
  ds^2 = - f(r) dt^2 + \frac{dr^2}{f(r)} \, . 
  \label{dS2ansatz}
\ee
In fact, by starting with the dS$_2$ solution in eq.~\eqref{dS2sol} and by applying the Weyl transformation-like field redefinition 
\be
g_{\mu\nu} \equiv \Omega(\phi)  \hat{g}_{\mu\nu} \, ,
\ee
one can always put the metric in the form of eq.~\eqref{dS2ansatz} by a coordinate change, as we will show soon. 
Here we always assume $\Omega(\phi)$ is regular. 
In what follows, we set $L=1$ for convenience. The $L$ scale can always be recovered by dimensional analysis.

Since $\phi=r$, we have $\Omega(\phi) = \Omega(r)$.  Starting from $f(r) = 1 -r^2$, 
the Weyl-transformed metric is given by 
\begin{align}
	ds^2=\Omega(r)\left(-f(r)dt^2+\frac{dr^2}{f(r)}\right)=-\Omega(r)f(r)dt^2+\frac{\Omega^2(r) dr^2}{\Omega(r) f(r)} \, .
	\label{Weylfieldred}
\end{align}
Defining the coordinate $\tilde{r}$ in terms of $r$ as
\begin{align}
	\tilde{r}(r) =\int^r \Omega(\hat{r}) d\hat{r}
	\label{tilder} \, ,
\end{align}
the line element can be written as
\begin{align}
ds^2=-\tilde{f}(\tilde{r})dt^2+\frac{d\tilde{r}^2}{\tilde{f}(\tilde{r})},\qquad \tilde{f}(\tilde{r})=\Omega(r(\tilde{r}))f(r(\tilde{r})) \, .
\label{gf}
\end{align}
One can check that the temperature of the solution remains unchanged upon Weyl transformation-like field redefinition because $t$ is unmodified. 
In this way, we can always make the metric in the form eq.~\eqref{dS2ansatz}. However, 
note that the dilaton is no more linear in terms of the new radial coordinate $\tilde{r}$ 
\be
\phi =  r(\tilde{r}) \,.
\ee 
In the end, this is just a field-redefinition, so physics should not be modified. Since the vacuum of JT gravity is characterized by both the metric and the dilaton, reasonable physical quantities should be determined by taking into account both.  While the on-shell action is invariant under such field-redefinitions, the volume is not, because it is determined by the metric only. In other words, action complexity is invariant, but the volume (precisely, geodesic length) changes, as we will see explicitly.

Given the metric ansatz in eq.~\eqref{dS2ansatz}, we now present the procedure to compute the length of geodesic following the approach of \cite{Jorstad:2022mls,Chapman:2021eyy}. As in the action computation, 
we focus on the portion of spacetime with $0 \leq r < \infty$.

In EF coordinates \eqref{EF}, the line element  on the right side of the Penrose diagram covered by ${(v_R,r)}$ or ${(u_R,r)}$ coordinates is
\begin{align}
	ds^2=-f(r)dv_R^2+2dv_Rdr=-f(r)du_R^2-2du_Rdr \, .
	\label{dsinef}
\end{align}
The length of geodesics in this geometry reads
\begin{equation}
	V = \int ds \sqrt{-f \dot{v}_R^2+2 \dot{v}_R \dot{r}} = \int ds \sqrt{-f \dot{u}_R^2-2 \dot{u}_R \dot{r}} \, ,
	\label{volume2}
\end{equation}
where the dot denotes the derivative with respect to the geodesic parameter $s$. 
Since the geodesic is spacelike, we can always choose a parameterization such that the integrand in eq.~\eqref{volume2} is unity: 
\be
\begin{aligned}
	&-f \dot{v}_R^2+2 \dot{v}_R \dot{r}=1 \quad \to \quad \dot{r} = \frac{1+ f \dot{v}_R^2}{2 \dot{v}_R} \, , \\
	&-f \dot{u}_R^2-2 \dot{u}_R \dot{r}=1 \quad \to \quad \dot{r} = -\frac{1+ f \dot{u}_R^2}{2 \dot{u}_R} \, .
	\end{aligned}
\end{equation}
The equations of motion obtained from eq.~\eqref{volume2} by varying $u_R,v_R$ lead to the conserved quantity
\be
\begin{aligned}
	P = \frac{\delta V}{\delta \dot{v}_R} = -f \dot{v}_R + \dot{r} = \frac{1- f \dot{v}_R^2}{2 \dot{v}_R} \, , \\
	P=\frac{\delta V}{\delta \dot{u}_R}=-f \dot{u}_R - \dot{r} = \frac{1- f \dot{u}_R^2}{2 \dot{u}_R} \, .
\end{aligned}
\label{Peq}
\ee
We can thus express $\dot{u}_R, \dot{v}_R,$ and $\dot{r}$ in terms of the conserved quantity $P$ as
\begin{equation}
    \dot{r}_{\pm} =\pm\sqrt{f+P^2} \, , \qquad
	\dot{u}_{R \pm}=\frac{-P - \dot{r}_{\pm}}{f} \, , \qquad 
	\dot{v}_{R \pm}=\frac{-P + \dot{r}_{\pm}}{f} \, .	
	\label{dotv}
\end{equation}
The  sign $\pm$ in the above equation indicates whether the parameter $s$ increases or decreases along the direction of  increasing $r$.
 There is a turning point where $f(r_t)+P^2=0$. 
{ We consider geodesics anchored at the left and right stretched horizons $r =r_{st}$, defined in eq.~\eqref{stretched}. We take by convention the parameter $s$ to increase from the left stretched horizon to the right one.}
In the chosen gauge, the volume is given by 
\begin{equation}
	V(P)= \int ds = 2 \int_{r_{st}}^{r_{t}} \frac{dr}{\dot{r}_+} = 2 \int_{r_{st}}^{r_{t}} \frac{dr}{\sqrt{f(r)+P^2}} \, ,
	\label{volume}
\end{equation}
where by symmetry we integrate over the left half of the geodesic and we introduce a multiplicative factor of $2$. 
The dependence of volume on the stretched horizon time $t_R = - t_L = \tau$ is encoded in the conserved quantity $P$.
{ For $P>0$, geodesics explore the region beyond the future cosmological horizon \cite{Chapman:2021eyy,Jorstad:2022mls}. In this case, the right portion of geodesics is fully covered by the EF coordinate $u_R$, while the left portion is fully described by $v_L$, see figure \ref{fig:penrose-dS}.} 
The time dependence of $P$ can thus be computed by
\begin{align}
	u_R(r_{st})- u_R(r_t) & =  \int_{r_t}^{r_{st}} \frac{\dot{u}_{R-}}{ \dot{r}_-} dr =  \int_{r_{st}}^{r_t}  T \le P,r \ri dr  \, , \label{v_eq} \\
	v_L (r_t) - v_L (r_{st}) & =   \int_{r_{st}}^{r_t} \frac{\dot{v}_{L+}}{ \dot{r}_+}  dr =  \int_{r_{st}}^{r_{t}} T \le P,r \ri dr \, ,
	\label{bdytime}
\end{align}
where 
\begin{align}
	T \le P,r \ri \equiv \frac{\sqrt{f(r)+P^2}-P}{f(r) \sqrt{f(r) + P^2}} \, .
\end{align}
Summing up eqs. \eqref{v_eq} and \eqref{bdytime} and using the definition of the EF coordinates, we end up with
\begin{equation}
	\tau(P) = r^{\ast} (r_{st})  - r^{\ast}_t + \int_{r_{st}}^{r_t} T (P,r) dr \,= -\int_{r_{st}}^{r_{t}} \frac{P}{f(r)\sqrt{f(r)+P^2}} dr \, ,
	\label{time_int}
\end{equation}
where we have defined $r^{\ast}_t  \equiv r^{\ast} (r_t)$.{ Note that the case $P<0$ corresponds to a time-reflection $\tau \to -\tau$.}

As we have seen, the metric resulting from Weyl field-redefinition is given by eq.~\eqref{Weylfieldred}. So, eqs.~\eqref{volume} and \eqref{time_int} still hold true, provided that we replace $r,f(r) \rightarrow \tilde{r},\tilde{f}(\tilde{r})$ given by eqs.~\eqref{tilder} and \eqref{gf}. Therefore, after the Weyl field-redefinition, the volume and the stretched horizon time are given by 
\begin{align}
	&V(P) = 2 \int_{\tilde{r}_{st}}^{\tilde{r}_{t}} \frac{d\tilde{r}}{\sqrt{\tilde{f}(\tilde{r})+P^2}}= 2 \int_{r_{st}}^{r_{t}} \frac{\Omega(r)}{\sqrt{\Omega(r)f(r)+P^2}} dr \, ,
	\label{genvptpfom1} \\
	&\tau(P) = \tilde{r}^{\ast} (\tilde{r}_{st}) - \tilde{r}^{\ast}_t  + \int_{\tilde{r}_{st}}^{\tilde{r}_{t}} \tilde{T} (P,\tilde{r}) d\tilde{r} \,= - \int_{r_{st}}^{r_{t}}\frac{P }{f(r)\sqrt{\Omega(r)f(r)+P^2}} dr \, ,
	\label{genvptpfom2}
\end{align}
where $\tilde{r}_t$  is the turning point obtained by 
\begin{align}
	\tilde{f}(\tilde{r}_t)+P^2=0 \, ,
	\label{rttup}
\end{align}
and $\tilde{r}^*$ is the tortoise coordinate defined by
\begin{align}
	\tilde{r}^*(\tilde{r}) = \int^{\tilde{r}} \frac{d\hat{r}}{\tilde{f}(\hat{r})} \, .
	\label{tilr}
\end{align}
To illustrate the Weyl dependence of the volume, we consider $\Omega(\phi) = \Omega(r)$ of the form 
\begin{align}
	\Omega(\phi)=  \phi^w  =  r^w \, ,
	\label{gr}
\end{align}
so that we get
\begin{align}
	\tilde{r}={r^{w+1}\over w+1} \, . 
	\label{tilrx} 
\end{align}
By choosing a few specific values of $w$,
we now show in detail how the volume in dS spacetime changes as we change $w$.

\subsubsection{$w=0$}
\label{w=0}

We consider the dS blackening factor given by eq.~\eqref{dsbf} with $L=1$. Thus, the cosmological horizon is at $r=1$. 
The stretched horizon is specified by the location where the dilaton takes the fixed value 
\begin{align}
	\phi_{st} = r_{st} = \rho \, .
	\label{strecbdds}
\end{align}
The turning point of the geodesic elongating between the right and left stretched horizons is
\begin{align}
	r_t = \sqrt{1+P^2} \, .
	\label{rttp}
\end{align}
The volume and the boundary time can be computed explicitly and are given by 
\begin{align}
	V(P)&=2\int_{\rho}^{\sqrt{1+P^2}} \frac{dr}{\sqrt{f(r)+P^2}}=\pi -  2 \, {\rm arctan}\left[ \frac{\rho}{\sqrt{P^2+1-\rho^2}}\right] \, , \nonumber\\
	\tau(P)&=-\int_{\rho}^{\sqrt{1+P^2}} \frac{P}{f(r)\sqrt{f(r)+P^2}} dr  ={\rm arctanh}\left[\frac{P\rho}{\sqrt{P^2+1-\rho^2}}\right] \, ,
	\label{vtinx1ds}
\end{align}
as found in \citep{Jorstad:2022mls}.
If we set $\rho=0$, geodesics stretch between the poles of dS spacetime. In this case, we get $V(P)=\pi$ and $\tau(P)=0$, which is consistent with the result of \cite{Chapman:2021eyy}.\\


\subsubsection{$w=2$}
\label{w=2}

For the case $w=2$, we have 
\begin{align}
	\tilde{r}={r^{3}\over 3} \, ,
	\label{rt}
\end{align}
so the blackening factor becomes 
\begin{align}
	\tilde{f}(\tilde{r})={(3 \tilde{r})}^{2/3}(1-{(3\tilde{r})}^{2/3}) \, .
	\label{ftdsx2}
\end{align}
The turning point in terms of the original coordinate $r$ is given by 
\begin{align}
	r_t=\frac{1}{\sqrt{2}} \sqrt{1+\sqrt{1 + 4 P^2}} \, .
	\label{rtdsx2}
\end{align}

\begin{figure}[htbp!]
	\centering
	\includegraphics[keepaspectratio,scale=0.8]{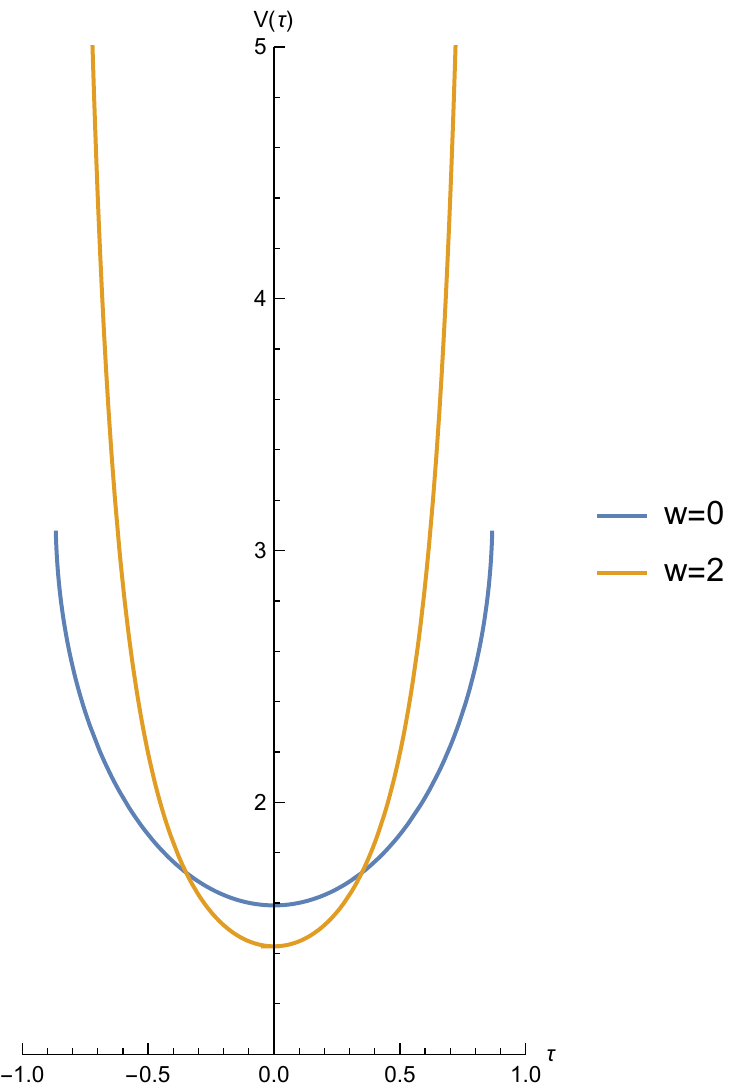}
	\caption{Plot of $V(\tau)$ for fixed values of $w$.  As $\tau \to \tau_{\infty}$, the refined volume behaves differently depending on $w$. The $w=0$ result is bounded by $\pi$. Instead, the refined volume  for $w=2$ diverges, due to the additional Weyl (dilaton) contribution from the point near the future infinity. We set $\rho=0.7$ and $\tau_{\infty}={\rm arctanh} (\rho) = 0.88$. }
	\label{fig1}
\end{figure}

Since the Weyl transformation only changes the metric, and not the dilaton, the location of the stretched horizon is still given by eq.~\eqref{stretched}. The expressions for the volume and the boundary time are given by 
\begin{align}
	&V(P)=2 \int_{\tilde{r}_{st}}^{\tilde{r}_t} \frac{d\tilde{r}}{\sqrt{\tilde{f}(\tilde{r})+P^2}}=2\int_{\rho}^{r_t} \frac{r^2}{\sqrt{r^2(1-r^2)+P^2}} dr \, ,\nonumber\\
	&\tau(P)=-\int_{\tilde{r}_{st}}^{\tilde{r}_t}  \frac{P}{\tilde{f}(\tilde{r})\sqrt{\tilde{f}(\tilde{r})+P^2}} d\tilde{r} = -\int_{\rho}^{r_t} \frac{P}{(1-r^2)\sqrt{r^2(1-r^2)+P^2}} dr \, .
	\label{vtinx2ds}
\end{align}
In order to compare the cases $w=0,2$ described above, in figure \ref{fig1} we present some plots of the volume as a function of time on the stretched horizon. 

Remarkably, for $w=2$ the volume diverges at the critical time, qualitatively matching the behavior of action complexity. Below, we consider refined volume for complexity.

\subsection{Refined volume for complexity}
\label{rvforc}

As we have seen, despite a field redefinition should not influence physics, volume complexity is clearly modified by this kind of transformation. Moreover, there is a discrepancy between volume complexity and action complexity in JT gravity.  In this subsection, we look for a Weyl factor of the form in eq.~\eqref{gr} with an appropriate parameter $w$ for which the transformed volume complexity behaves qualitatively as action complexity. We refer to the volume with this specific choice of $w$ as refined volume. 

This can be easily done as follows. 
We recall that the action complexity we computed in section \ref{cainjt} is obtained by dimensional reduction from dS$_3$. Therefore, action complexity in dS$_2$ JT gravity is essentially the same as in dS$_3$ Einstein gravity. For Einstein gravity in dS$_{d+1}$, both CV and CA conjectures are studied in detail in \cite{Jorstad:2022mls}, where the authors have found
\be
\label{Vtintegral}
V \sim 
\int^{r_t}_{\rho} dr \, \frac{(r/L)^{2(d-1)} }{\sqrt{P^2+f(r)\,\,(r/L)^{2(d-1)}}} \, , \qquad
\tau \sim 
\int^{r_t}_{\rho} \frac{dr}{L}\, \frac{- P}{f(r)\sqrt{P^2+f(r)\,\,(r/L)^{2(d-1)}}}  \, , 
\ee
see eq.~(5.9) and (5.11) in \cite{Jorstad:2022mls}.
In dS$_{d+1}$ ($d \ge 2$), both volume and action complexity grow as $r_+^{d-2}$. Therefore, a direct comparison of the above result with our eqs.~\eqref{genvptpfom1} and \eqref{genvptpfom2} for dS$_3$ determines
\be
\label{ds2wef}
{\Omega(\phi) = \phi^{2} \, .}  
\ee
With this choice, corresponding to $w = 2$ in eq.~\eqref{gr},
the refined volume matches volume complexity in Einstein gravity dS$_{3}$, which reduces to JT gravity dS$_2$. 
In this way, in JT gravity one can define the Weyl transformed refined volume which behaves as action complexity.\footnote{
More generally, to get the volume complexity result of conventional JT gravity to match with the corresponding result of Einstein gravity in $dS_{d+1}$, the Weyl factor should be taken to be
$\Omega(\phi)=\phi^{2 (d-1)}$.
}

We end this section by pointing out that, in a similar spirit as the refined volume we have discussed, modifications of the CV conjecture in non-Einstein gravity theories with higher derivative terms have been proposed in \cite{Alishahiha:2015rta,Bueno:2016gnv,Alishahiha:2017hwg,Hernandez:2020nem}.

\section{Summary and Discussion}
\label{disc}

In this paper, we have confirmed that action complexity in dS$_2$ JT gravity, with the effect of the dilaton properly taken into account, diverges at the finite time $\tau = \tau_{\infty}$ given in eq.~\eqref{tau-infty}. Instead, volume complexity reaches an $\mathcal{O}(1)$ critical value, although with a diverging rate as the critical time $\tau=\tau_{\infty}$ is approached. We point out that for larger times $\tau$, when geodesics anchored at the stretched horizon meet the future infinity of dS$_2$, the definition of volume complexity requires particular care. An analysis has been done in \cite{Jorstad:2022mls}, where extremal curves are replaced by piecewise geodesics including a cutoff portion nearby the future infinity. With this prescription, volume is divergent too, but at the edge of the two regimes, there is an infinite jump. On the other hand, action complexity is  well-defined even after the critical time $\tau_\infty$, when the WDW patch is cut off by the future infinity.
 The crucial difference between volume and action complexity in two dimensions is due to the dilaton. Since the solution of JT gravity is characterized by both the metric and the dilaton, the effect of the latter is crucial. The dilaton naturally arises by dimensional reduction of three-dimensional dS spacetime, which provides the ``right action'' in eq.~\eqref{tot_action}. This is because the dimensional reduction of dS$_3$ yields dS$_2$ with a linear dilaton, which is the solution of JT gravity. Consistently, by evaluating the on-shell action in eq.~\eqref{tot_action}, we obtain the same action complexity as for dS$_3$ \cite{Jorstad:2022mls}, see eq.~\eqref{finalCA}. 
Note that all contributions in eq.~\eqref{tot_action} are boundary terms. This is  expected since the bulk JT gravity action vanishes on-shell. We also stress that in dS spacetime the dilaton diverges at the future infinity. This behaviour of the dilaton is crucial for action complexity in JT gravity to show the hyperfast growth predicted in \cite{Susskind:2021esx}. 

We have observed that the CV conjecture suffers a Weyl field-redefinition issue. A field-redefinition should not affect the physics, but volume complexity is manifestly modified by Weyl field-redefinition. In higher dimensions, there is a preferred frame, the Einstein frame. However, in two dimensions, since $\sqrt{g}R$ is invariant under Weyl transformation, there is no such preferred frame and this leads to an ambiguity in the definition of the CV conjecture. On the contrary, the CA conjecture is free from this problem, because the on-shell action is invariant under field-redefinition. As a way out, we have proposed a particular gauge choice in dS$_2$ JT gravity under which the so-called refined volume behaves qualitatively as action complexity. The full analysis of refined volume in JT gravity is an open issue on which we hope to come back in the near future.

Quantum computational complexity quantifies how hard it is to build a target state from a given reference one in terms of the minimum number of elementary gates required. The hyperfast growth of action complexity, as well as its divergence at finite time, indicates that the late time growth of dS spacetime is too difficult to capture, as conjectured in \cite{Susskind:2021esx}. 
Susskind conjectured that the boundary dynamics of JT gravity in dS$_2$  is related to the double-scaled SYK model. It would be quite interesting to investigate this particular limit of SYK and to evaluate complexity  directly in such a setting.

Another interesting possible direction is to embed dS$_2$ in asymptotic AdS$_2$ \cite{Anninos:2017hhn,Anninos:2018svg}. Previous studies of holographic complexity in this kind of geometries have been done using the CV conjecture \cite{Chapman:2021eyy}. It would be interesting to investigate how the results are modified by using the CA conjecture. It also remains to be seen how the refined volume conjecture in these more general kinds of spacetimes works.

\acknowledgments
We would like to thank Stefano Baiguera, Shira Chapman, and Shan-Ming Ruan for their helpful comments on our draft. The work of TA and NI was supported in part by JSPS KAKENHI Grant Number 21J20906(TA), 18K03619(NI). The work of NI, SS, and NZ was also supported by MEXT KAKENHI Grant-in-Aid for Transformative Research Areas A “Extreme Universe” No. 21H05184.

\bibliographystyle{JHEP}

\bibliography{bibliography}

\end{document}